\documentclass[10pt]{article}
\usepackage{amssymb,units,amsmath}
\usepackage{bbold} 
\usepackage{euler,newcent}
\usepackage{graphicx}
\usepackage{dcolumn}
\usepackage{bm,bbm,mathtools,esvect}
\usepackage{slashed,fullpage}
\usepackage{tgbonum}
\title{ {\sc Winding number and homotopy \\ for quaternionic curves}\\ \vspace{0mm}} 
\author{ {\tt SERGIO GIARDINO}\footnote{\tt sergio.giardino@ufrgs.br}\\ 
\vspace{0cm}\\
{\small\it Departamento de Matem\'atica Pura e Aplicada }\\ 
{\small\em Universidade Federal do Rio Grande do Sul (UFRGS)}\\ 
{\small\em  Caixa Postal 15080, 91501-970 Porto Alegre RS}\\
{\small\em  Brazil}}

\begin{document}
\date{} 

\newtheorem{theorem}{Theorem}[section] 
\newtheorem{remark}{Remark}[section] 
\newtheorem{lemma}{Lemma}[section] 
\newtheorem{proposition}{Proposition}[section] 
\newtheorem{corollary}{Corollary}[section] 
\newtheorem{definition}{Definition}[section]

\maketitle

\begin{abstract} \noindent Following a recent approach to quaternionic curves, 
we defined the quaternionic  polar angle that enabled us to define global properties of quaternionic curves,
namely the  winding number and the homotopy concept. The results admit various applications, including further analogies 
to plane curves, and physical applications.

\vspace{2mm}

\noindent{\bf keywords:} curves in Euclidian space; Hyper-K\"ahler and quaternionic K\"ahler geometry; Quaternion and other division algebras.

\vspace{1mm}

\noindent{\bf MSC class. codes:} 53A04; 53C26; 11R52.
\end{abstract}

\tableofcontents

\section{INTRODUCTION} 
The study of quaternionic curves is a relatively novel area of research, beginning in \cite{Bharathi:1987kbn,Gunes:1994mmt} and taking impulse in recent times with several interesting results, including theoretical developments \cite{Aksoyak:2019}, applications of various sorts of quaternionic curves \cite{Coken:2004otq,Hacisalihoglu:2011gok,Gungor:2011scq,Ilarslan:2013bma,EJPAM1979,Bektas:2016qoc,Kisi:2017aap,Coken:2017nqc,Karadag:2019scq,Kizilay:2020eqc,Kahraman:2020dqn}, and the study of evolvents and evolutes \cite{Gungor:2013qie,Onder:2020gqi,Altun:2017scy,Senyurt:2017scy,Hou:2018gia}. More recently, a novel approach \cite{Giardino:2021onv} took benefit from the algebraic properties of quaternions to obtain the Frenet-Serret equations, and to determine evolutes and evolvents of quaternionic curves.
In this article, the formalism introduced in \cite{Giardino:2021onv} is applied to define the winding numbers of quaternionic curves and to determine several properties, and thus establishing an analogy to the well known subject of  curves in $\mathbbm R^2$ \cite{Hopf:1935dtk}, which deserves a section in current textbooks on differential geometry \cite{Hilario:2003dcg,Montiel:2009ras,Carmo:2016rgc}. 

More specifically, this article reaches several targets. The first one is to generalize the winding number of plane curves in $\mathbbm R^2$ or $\mathbbm C$ to quaternionic curves, and to apply it  to obtain a quaternionic version for the fundamental theorem of algebra. Despite the previous quaternionic versions of the fundamental theorem of algebra \cite{Niven:1944fta,Zhang;1997qmq},  to the best of our knowledge it was  not interpreted in terms of winding numbers, and this novel approach is given here. However, the deepest insight that emerges is a method to generalize the complex curves through the transformation of the complex imaginary unit $\,i\,$ into the pure imaginary quaternionic function $\,\omega(t),\,$ that is a quaternionic curve parametrized  by the real variable $\,t.\,$ This simple, although not obvious idea, is what proportionate the novel view of quaternionic curves presented in this article. 

The notational conventions and the elementary facts about quaternioninc hyper-complexes that will be useful in this article are available in the Section \ref{A}, while the main subject of the article, and the novel results are comprised within Sections \ref{RN} and \ref{H}.

\section{QUATERNIONIC BACKGROUND\label{A} } 
There are many sources that introduce and describe the quaternions \cite{Ward:1997qcn,Garling:2011zz,Rocha:2013qtt,Morais:2014rqc}, and we only describe the notation for representing quaternions that will be useful in this article, and of course nothing in this section is new.
Thus, we initially state that quaternionic numbers ($\mathbbm H$) are hyper-complexes that, whenever $\,q\in\mathbbm H$, then 
\begin{equation}\label{hn01} 
q=x_0+x_1i+x_2 j+x_3k,\qquad\mbox{where}\qquad x_\mu\in\mathbbm R,\qquad\mbox{and}\qquad\mu\in\{0,\,1,\,2,\,3\}. 
\end{equation} 
The anti-commuting imaginary units $\,i,\,j\,$ and $\,k\,$  comply with 
\begin{equation}\label{hn02} 
i^2\,=\,j^2\,=\,k^2\,=-1,\qquad\mbox{and}\qquad ijk=-1. 
\end{equation} 
Likewise the complex numbers, the quaternionic conjugate and the quaternionic norm read
\begin{equation}\label{hn03} 
\overline q\,=\,x_0-x_1i-x_2 j-x_3k,\qquad\qquad |q|\,=\,\sqrt{q\overline q\,}\,=\,\sqrt{x_0^2+x_1^2+x_2^2+x_3^2\,}. 
\end{equation} 
Adolf Hurwitz proved that quaternions comprise one of the four division algebras \cite{Hurwitz:1898hqt}, and the other ones are the reals ($\mathbbm R$), the complexes ($\mathbbm C$) and the octonions ($\mathbbm O$).  The extended or cartesian notation for quaternions (\ref{hn01}) may
also be switched to
\begin{equation}\label{hn04} 
q=x_0+\omega |\bm x|\qquad \mbox{where}\qquad \omega=\frac{\bm x}{|\bm x|}\qquad \mbox{and}\qquad x=x_1i+x_2 j+x_3k. 
\end{equation} 
This representation encompasses the composed imaginary unit $\omega$, so that $\omega^2=-1$. Moreover, $x_0$ and $\bm x$ are respectively nominated the scalar (or temporal) and the vector (or spatial) parts of the quaternion. This notation has the opportune feature that the scalar and the vector components are commutative, and the inconvenience that every quaternionic number has its proper imaginary unit $\omega$ that neither commutes nor anti-commutes with the imaginary unit of other quaternionic number. The polar notation of the cartesian quaternion is such that
\begin{equation}\label{hn05} 
q\,=\,|q|\big(cos\theta+\sin\theta\,\omega\big),\qquad\mbox{where}\qquad\theta\in[0,\,\pi],
\end{equation} 
and the imaginary component of (\ref{hn04}) is always positive, and hence the range of the polar angle $\,\theta\,$ is narrow when compared to the complex numbers. Ultimately, the symplectic notation for quaternions asserts that
\begin{equation}
\label{hn06} q=z_0+z_1 j,\qquad\mbox{where}\qquad z_0=x_0+x_1 i\qquad\mbox{and}\qquad z_1=x_2 +x_3 i. 
\end{equation} 
The symplectic notation is besides not unique, and (\ref{hn06}) is exchangeable with 
\begin{equation}\label{hn07} 
q=z_0+\overline\zeta\, k \qquad\mbox{where}\qquad \zeta=x_3 +x_2 i, 
\end{equation} 
and two supplementary possibilities replacing $i$ with $j$ and $k$ in (\ref{hn06}). The  polar form displayed in symplectic notation renders
\begin{equation}\label{hn08} 
q\,=\,|q|\Big(\cos\vartheta e^{i\phi}+\sin\vartheta e^{i\psi}j\Big)\qquad\mbox{where}\qquad \vartheta\in\left[0,\frac{\pi}{2}\right]\qquad\mbox{and} \qquad \phi,\,\psi\in[0,\,2\pi]. 
\end{equation}
The range of the symplectic polar angle $\,\vartheta\,$ is even more restrictive than (\ref{hn05}) thanks to the positive definite character of their trigonometric functions, that are defined in terms of  the moduli of the complex components. From reference \cite{Harvey:1990sca}, we adopt the scalar product for quaternions 
\begin{equation}\label{hn10} 
\big\langle p,\,q\big\rangle\,=\,\mathfrak{Re}\big[\,p\overline q\,\big], 
\end{equation} 
which authorzes us to state that if $p$ and $q$ are orthogonal, then 
\begin{equation}\label{hn11} 
\big\langle p,\,q\big\rangle\,=\,0, 
\end{equation} 
and also that $p$ and $q$ are parallel whenever
\begin{equation}\label{hn12} 
\big\langle p,\,q\big\rangle\,=\,\,p\overline q.
\end{equation} 
Perceivable, properties (\ref{hn10}-\ref{hn12}) are valid also for complex numbers, and can be understood as generalizations. The inner product (\ref{hn10}) enables us to gain the orthogonality relations, which in cartesian notation are 
\begin{equation}\label{hn13} 
\big\langle q,\,e_\ell q\big\rangle = 0,\qquad\mbox{where}\qquad e_\ell=\big\{i,\,j,\,k\big\}. 
\end{equation} 
Quaternions undoubtedly comprise a four dimensional real vector space, but the polar notations (\ref{hn05}) and (\ref{hn06}) subsequently give
\begin{equation}\label{hn14} 
\langle q,\,\omega q\rangle =0,\qquad\mbox{and}\qquad\langle q,\, q j\rangle =0,
\end{equation} 
where the quaternions comprise  two dimensional real vector spaces in terms of these notations. An unexpected property, considering that the space has actually four real dimensions from (\ref{hn13}). To illuminate this point, let us remember that the restriction of the range of $\theta$ in (\ref{hn05}) and $\vartheta$ in (\ref{hn08}) indicates that negative signals ought to be assimilated in the complex structure in order to keep the polar angles within the correct range. In the simplest case, the negative of a cartesian polar quaternion $\,q$, we have 
\begin{equation}\label{hn15} 
-q=(-1)(\cos\theta+\omega\sin\theta).
\end{equation} 
If $\,-q\,$ were complex, the product would signify a rotation of the polar angle, so that $\theta\to\theta+\pi$, something not allowed in the quaternionic case. Hence, the negative signal have to be accommodated within the imaginary unit $\omega$, and the precise rotation of the angle will be $\theta\to\pi-\theta$. Furthermore, this multiplication reverses the orientation of the polar angle, and hence the quaternionic polar angle is not orientable accoding to this operation. A desiring outcome in dimensions higher than two, where the polar angle is supposed to be non-orientable. Using the above precept, we further have
\begin{equation}\label{hn16} 
\left. 
\begin{array}{l} \quad	\;\, q\,=\,\cos\theta+\omega\sin\theta\\ \\ \;\;\omega q\,=\,\cos\left(\theta+\frac{\pi}{2}\right)+\omega\sin\left(\theta+\frac{\pi}{2}\right)\\ \\ \omega^2 q\,=\,\cos\left(\pi-\theta\right)+\widetilde\omega\sin\left(\pi-\theta\right)\\ \\ \omega^3 q\,=\,\cos\left(\frac{\pi}{2}-\theta\right)+\widetilde\omega\sin\left(\frac{\pi}{2}-\theta\right)\\ 
\end{array} \right\} 
\qquad\theta\in \left[0,\frac{\pi}{2}\right]\qquad\mbox{and}\qquad \widetilde\omega=-\omega.
\end{equation} 
The $\,\theta\in\left[\frac{\pi}{2},\,\pi\right]\,$ case can also be readily obtained, and the conditions of (\ref{hn05}) are hence satisfied for all the  possible cases. We can also apprehend that there are two orthogonal pairs $(q,\,\omega q)$ and $(\omega^2 q,\,\omega^3 q)$, thus counting the four dimensions of the space, which are not equivalent because their imaginary unit is different. Adopting these ideas, we generally have 

\begin{equation}\label{hn17} 
q(\theta_1+\theta_2)\,=\,\left\{ \begin{array}{ll} \cos\theta_0+\omega\sin\theta_0 & n\quad\mbox{even}\qquad \theta_0\in\left[0,\,\pi\right]\\ \\ \cos\left(\pi-\theta_0\right)+\widetilde\omega\sin\left(\pi-\theta_0\right)& n\quad\mbox{odd,}\qquad \theta_0\in\left[0,\,\frac{\pi}{2}\right]\\ \\ \cos\left(\frac{\pi}{2}-\theta_0\right)+\widetilde\omega\sin\left(\frac{\pi}{2}-\theta_0\right) & n\quad\mbox{odd,}\qquad \theta_0\in\left[\frac{\pi}{2},\,\pi\right]\\ \\ \quad\;\,\theta_1+\theta_2\,=\,n\pi+\theta_0\qquad \mbox{and}\qquad \widetilde\omega=-\omega& \end{array} \right. 
\end{equation} 
Replicating this procedure using the symplectic notation (\ref{hn08}), we obtain
\begin{equation}\label{hn18} 
\left. \begin{array}{l} \;\;\; q\, =\,\cos\vartheta e^{i\phi}+\sin\vartheta e^{i\psi}j\\ \\ \;\, q j\,=\,\cos\left(\frac{\pi}{2}-\vartheta\right)e^{i(\psi-\pi)}+\sin\left(\frac{\pi}{2}-\vartheta\right)e^{i\phi}j\\ \\ q j^2\,=\,\cos\vartheta e^{i(\phi-\pi)}+\sin\vartheta e^{i(\psi-\pi)}j\\ \\ q j^3\,=\,\cos\vartheta\left(\frac{\pi}{2}-\vartheta\right)e^{i\psi}+\sin\left(\frac{\pi}{2}-\vartheta\right)e^{i(\phi-\pi)}j\\ 
\end{array} \right\} \qquad\vartheta\in \left[0,\frac{\pi}{2}\right] 
\end{equation} 
and also 
\begin{equation}\label{hn19} 
q(\vartheta_1+\vartheta_2)\,=\, \left\{ 
\begin{array}{ll} \cos\vartheta_0 e^{i\phi}+\sin\vartheta_0 e^{i\psi}j & \quad n=0\;\mbox{mod}\;4\\ \\ \cos\left(\frac{\pi}{2}-\vartheta_0\right)e^{i(\psi-\pi)}+\sin\left(\frac{\pi}{2}-\vartheta_0\right)e^{i\phi}j & \quad n=1\;\mbox{mod}\;4\\ \\ \cos\vartheta_0 e^{i(\psi-\pi)}+\sin\vartheta_0 e^{i(\phi-\pi)}j &\quad n=2\;\mbox{mod}\;4\\ \\ \cos\vartheta\left(\frac{\pi}{2}-\vartheta_0\right)e^{i\psi}+\sin\left(\frac{\pi}{2}-\vartheta_0\right)e^{i(\phi-\pi)}j&\quad n=3\;\mbox{mod}\;4 
\end{array} \right. 
\end{equation} where \[ \vartheta_1+\vartheta_2=\vartheta_0 + n\frac{\pi}{2},\qquad\vartheta_0\in\left[0,\,\frac{\pi}{2}\right] \qquad\mbox{and}\qquad n\in\mathbbm N. \] 
Therefore the set of necessary quaternionic concepts is complete, comprising the polar notations, the inner product, and the requiriments for orthogonality.

\section{ANGULAR FUNCTION AND WINDING NUMBER\label{RN}} 
We separate the analysis according to the quaternionic notations defined in the Section \ref{A}: the polar cartesian notation (\ref{hn04}-\ref{hn05}) and the symplectic notation (\ref{hn07}-\ref{hn08}) for quaternion curves, and therefore  the global property of winding number that is defined to plane curves in $\mathbbm R^2$ \cite{Hilario:2003dcg} will be extended to curves parametrized in  quaternionic terms.
\subsection{Polar cartesian quaternionic curves} 
The parametrization of quaternionic curves in the $\,\mathbbm H\equiv\mathbbm R\times \mathbbm R\omega\,$ space, where (\ref{hn04}-\ref{hn05}) hold, is similar to the complex description of plane curves in the $\,\mathbbm R\times \mathbbm R i\,$ plane, but the quaternionic case can be considered as a half space, and the imaginary unit will be generalized to a function $\,\omega:I\to\mathbbm H,\,$ where we define to the rest of the paper the real interval
\begin{equation}
I=[a,\,b],\quad\mbox{and their variable}\quad t\in I.
\end{equation} 
The concept of quaternionic curve is analogous to regular real plane curves either in $\mathbbm R^2$ or in $\,\mathbbm C,\,$ so that

\begin{definition}[Regular quaternionic curves] The parametrized curve $q:I\to\mathbbm H$ is regular in $\,I\subset\mathbbm R\,$ if $\;|q'(t)|\neq 0\quad\forall t\in I.$
\end{definition}

Therefore, we can prove that
\begin{proposition}[ Imaginary quaternionic function ]\label{rnP00} Let $x=x(t)$ be a pure imaginary regular quaternionic curve. Therefore,
\[
\begin{aligned}
i. &\qquad\overline{\omega}'=-\,\omega'\\
ii. &\qquad\omega\omega'=-\,\omega'\omega \qquad \mbox{where}\qquad \omega=\frac{x}{|x|}\\
iii.& \qquad\big(\omega'\big)^2=-\frac{1}{|x|^2}\Big(|x|'-|x'|\Big)^2\\
iv. & \qquad\omega\omega''+\omega''\omega=2|\omega'|^2 
\end{aligned}
\]
\end{proposition}
{\bf Proof:} The item (ii) comes immediately from the derivative of $\omega^2=-1$, and (ii) comes from
\[
\omega'=\frac{|x'|}{|x|}w-\frac{|x|'}{|x|}\omega,\qquad\mbox{where}\qquad w=\frac{x'}{|x'|}.
\]
where $\overline w=-\,w$. Item (iii) comes after squaring (iii) and using identity 
\[
\omega w+w\omega=-2\frac{|x|'}{|x'|}.
\]
Property (iv) is also immediate after deriving (i) and using (ii).

$\hfill\bm\square$

Proposition (\ref{rnP00}) establishes the relevant properties of the quaternionic imaginary unit in polar notation (\ref{hn04}).  Nonetheless, the definition of an angular function is also needed. In the case of a plane curve, an arbitrary point chosen out of their trace will be the vertex for the measured angle between two points of the curve. Keeping  this previous concept, a suitable expression for the angular function is as follows:

\begin{proposition}[The polar angular function]\label{rnP01} Let $q:I\to\,\mathbbm H\,$ be a regular quaternionic curve. The polar angular function of $q(t)$ around the regular curve $p_0(t)$ out of the trace of the curve $q$ is 
\begin{equation}
\theta(t)\,=\,\int_a^t\frac{\big\langle p(\tau)\omega(\tau),\,p'(\tau)\big\rangle}{\big| p(\tau)\big|^2}d\tau\qquad\quad \mbox{where}	\quad\qquad p(\tau)=q(\tau)-p_0(\tau), 
\end{equation} 
and $\omega(t)$ is the imaginary unit of the polar parametrization of $\,p(t)$.
\end{proposition} {\bf Proof:}
\noindent The parametrization 
\[ 
p(t)-p_0(t)=\rho(t)\Big(\cos\theta(t)+\sin\theta(t)\omega(t)\Big), 
\] 
leads to 
\[
 \big\langle (p-p_0)\omega,\,(p-p_0)'\big\rangle\,=\,\rho^2\theta' 
\] 
whose integration immediately gives the desired result. 

$\hfill\bm\square$

Interestingly, the imaginary unit $\omega$ does not affect the result.  
By way of example, let us consider the quaternion curve 
\begin{equation}\label{hn20} 
q(t)=R\,\omega(t)+R\Big[\cos t+\omega(t)\sin t\Big], \qquad t\in\big[0,\,2n\pi\big],\qquad n\in\mathbbm Z\backslash \{0\}.
\end{equation}
This curve can be seen as a sort of spiral whose projection over the $\mathbbm R\times\mathbbm R\omega$ plane is identical to a circumference of radius $R$ and center $R\omega$ for all values of the parameter $t$. Hence
\begin{equation}
p(t)=q(t)-p_0(t)\qquad\Rightarrow\qquad \langle p\omega,\,p'\rangle=R^2.
\end{equation}
To obtain the polar angle swept along the 	interval, and to keep  the positive character expected to the pure imaginary axis in the polar notation (\ref{hn05}), two separate integrals are needed. A first one for $t\in[0,\,\pi],$ where the imaginary direction is $\omega$, and a second integration for $t\in[\pi,\,2\pi]$. This second integral goes to the correct parametrization because $[\pi,\,2\pi]\to[\pi,\,0]$ and $\omega\to-\omega$. Consequently $\,\theta=2n\pi,\,$ after $n$ integrations, a perfect agreement to the complex plane after the transformation $\omega\to i$. After this example,	we define
\begin{definition}[The winding number]\label{rnD02} 
Let $q:I\to\,\mathbbm H\,$ be a continuous closed and piecewise differentiable quaternionic curve. The winding number of this curve along the continuous and differentiable $p_0(t):I\to\mathbbm H$ is  
\[	
\mathcal R(q,\,p_0)\,=\, \frac{1}{2\pi}\int_a^b\frac{\big\langle p(\tau)\omega,\,p'(\tau)\big\rangle}{\big| p(\tau)\big|^2}d\tau
\]
where
\[ 
	p(t)=q(t)-p_0(t),\qquad p(t)\in\mathbbm R\times\mathbbm R\omega,\qquad\mbox{and}\qquad \mathcal R(q,\,p_0)\in\mathbbm Z.
\] 
\end{definition}
$\hfill\bm\square$

The analogy between complex curves and quaternionic curves is thus completed. In the complex case a fixed point is chosen, while in the quaternionic case a continuous and smooth quaternionic curve is the reference to the winding number.  We observe that a variable radius of the curve does not change the result, as was seen in Proposition \ref{rnP01}. Moreover, we can prove that:

\begin{proposition}\label{rnP03} Let $q:I\to\mathbbm H\,$ be a closed regular curve. Therefore,  $\exists\, \rho>0$ such that, 
if $\, p(t):I\to\mathbbm H\,$ is a regular curve  that satisfies $|p(t)|\geq \rho$, then
\[
\mathcal R(p,\,q)=0.
\]
\end{proposition} 

\noindent{\bf Proof:} Let $\,\rho_0\,$ be the maximum value of $|q(t)|$, and $|p(\tau)|=\rho$ the minimum value of $|p(t)|$. If $\rho>\rho_0$, $\,p(\tau)\omega(\tau)\cap q(t)=\varnothing,$ and the polar angular function $\theta(t)$ defined by $q(t)-p(t)$ is such that $\theta(t)<\pi/2$, and hence $\theta(b)=0$.

$\hfill\bm\square$

\subsection{Symplectic quaternionic curves} 
The parametrization of quaternionic curves in the $\,\mathbbm C\times \mathbbm C j\,$ plane is different to the description of quaternionic curves parametrized in $\,\mathbbm R\times \mathbbm R \omega\,$ plane. If we naively try to repeat the proof of Proposition (\ref{rnP01}), defining 
\begin{equation} \label{hn200}
p(t)\,=\,\rho\Big(\cos\vartheta \,e^{i\phi}\,+\,\sin\vartheta \,e^{i\psi}\,j\,\Big),\qquad\mbox{where}\qquad t\in I=[a,\,b] 
\end{equation} 
and $\vartheta(t),\,\phi(t)\,$ and $\,\psi(t)\,$ are real functions, we get 
\begin{equation} \label{hn201}
\big\langle p j,\,p'\big\rangle\,=\,\rho^2\left[\,\vartheta'\cos\big(\phi-\psi\big)\,+\,\sin\vartheta\cos\vartheta\sin\big(\phi-\psi\big)\big(\phi+\psi\big)'\,\right], 
\end{equation} 
and we cannot obtain an integral expression ignoring the relation between the angular functions. This difference between the cases occurs because $\,p\omega\,$ and $\,p'\,$ are parallel quaternions, while $\,p j\,$ and $\,p'\,$ are not. Only parallel quaternions admit a simple angular polar function.

By way of example the quaternionic curve
\begin{equation}\label{hn21}
q(t)=R\Big(e^{i\phi}+e^{i\psi}j\Big)\,+\,R\Big[\cos\vartheta e^{i\phi}+\sin\vartheta e^{i\psi}j\Big],\qquad 
\vartheta\in\left[0,\,2n\pi\right]
\end{equation}
with smooth and continuous $\phi=\phi(t)$, and $\psi=\psi(t)$, and  $n\in\mathbbm Z$. In the particular case where
\begin{equation}
p_0(t)=R\Big(e^{i\phi}+e^{i\psi}j\Big),\qquad \phi=\psi,\qquad\mbox{and}\qquad p=q-p_0
\end{equation}
 $p=q-p_0,\,$ then 
\begin{equation}\label{hn22}
\big\langle p j,\,p'\big\rangle\,=\, R^2\vartheta'.
\end{equation}
As in the first case, we have to ascertain the parameters in order to have the correct integration, where $\vartheta\in[0,\,\pi/2]$. Hence, there are three changes to be done,
\begin{equation}
\begin{array}{lll}
i.&\left[\frac{\pi}{2},\,\pi\right]\to \left[\frac{\pi}{2},\,0\right],&\mbox{and}\qquad \phi\to\phi+\pi,\\ \\
ii.&\left[\pi,\,\frac{3\pi}{2}\right]\to \left[0,\,\frac{\pi}{2}\right],&\phi\to\phi+\pi\qquad\mbox{and}\qquad \psi\to\psi+\pi,\\ \\
iii.&\left[\frac{3\pi}{2},\,2\pi\right]\to \left[\frac{\pi}{2},\,0\right],&\mbox{and}\qquad \phi\to\phi+\pi.\\
\end{array}
\end{equation}
Consequently, the four integrals to be calculated are
\begin{equation}
\frac{1}{R^2}\int_0^{\pi/2} \big\langle p j,\,p'\big\rangle d\vartheta\,=\,\frac{\pi}{2}.
\end{equation}
Repeating the process $4n$ times, we obtain the total rotation number to be $2n\pi$, as expected.
 Definition \ref{rnD02} of the winding number also holds in the symplectic case, although choices such as $\psi=\phi$ have to be made.
\section{WINDING NUMBER AND HOMOTOPY\label{H}} 
As in the previous section, we keep the complex curves as the model for the quaternionic curves case. Thus, we define the parameter 
$\,\alpha\,$ of deformation of a quaternionic curve, so that
\begin{equation}
J\subset\mathbbm R,\qquad 0\in J\qquad\mbox{and}\qquad\alpha \in J.
\end{equation}
In analogy to complex curves:
\begin{definition}[Deformed curve] \label{HD01}
Let $\,q(t):I\to\mathbbm H\,$ be a continuous curve parametrized as $\,\mathbbm R\times\mathbbm R \omega.\,$ The regular curve $\,p(t,\,\alpha):I\times J\to\mathbbm H\,$ is a continuous deformation of $q$ if it satisfies
\[
p(t,\,0)\,=\,q(t).
\]
\end{definition}
As it is well known from plane curves, the angular function $\,\theta(t,\,\alpha)\,$ associated to the deformation is also continuous in the case of a fixed value $\alpha_0\in J$. The concept of continuous deformation permits the introduction of the spatial concept of homotopy, in analogy to the plane concept. In order to get it, let us then prove several necessary results.
\begin{lemma}\label{HP01} Let  
$\,q(\alpha,\,t):J\times I\to\mathbbm H\,$ be a continuous deformation of the regular curve $\,q(t):I\to\mathbbm H,\,$  and the regular curve $\,p(\alpha):J\to \mathbbm H\,$  is such that $\;p(\alpha,\,t)\cap q(\alpha,\,t)= \varnothing.\;$ The angular function $\,\theta(\alpha,\,t)\,$ associated to the points of $q(\alpha,\,t)$ using the point  $p(\alpha)$, such that $\,\theta(\alpha,\,a)=0,\,$ is a continuous function of $t$ and $\alpha$   
\end{lemma}
{\bf Proof:} Let us define the unit vector
\[
v_\alpha(t)=\frac{q(\alpha,\,t)-p(\alpha,\,t)}{\big|q(\alpha,\,t)-p(\alpha,\,t)\big|}
\]
Let us suppose that, given $r,\,s\in I$ we have that $\big\langle v_\alpha(t),\,v_\alpha(s)\big\rangle\neq -1.$  The continuity of $q(\alpha,\,t)$ permit us to built the sequences that obbey
\[
\big|\alpha_0-\alpha\big|<\frac{1}{n},\qquad\mbox{and}\qquad\big|t_n-s_n\big|<\frac{1}{n},\qquad\mbox{where}\qquad\alpha_0\in J\qquad
\mbox{and}\qquad n\in\mathbbm N.
\]
consequently, 
\[
\lim_{n\to\infty}\alpha_n=\alpha_0\qquad\mbox{and}\qquad\lim_{n\to\infty}s_n=\lim_{n\to\infty}t_n=t_0\in I
\]
Consequently, exists $\delta>0$ so that $|\alpha_0-\alpha|<\delta$ and $|s-t|<\delta$ so that $\big\langle v_\alpha(t),\,v_\alpha(s)\big\rangle\neq 1,$ a contradiction, and thus we can use the inner product build from the inner product is a continuous function. If we divide $I$ in segments such that $a=t_0<t_1<\dots t_n=b$, the have that $|t_{k+1}-t_k|<\delta$, and $|\alpha_0-\alpha|<\delta$ define the angular function as a sum of continuous functions over $k$, and thus $\theta(\alpha,\,t)$ is continuous.

$\hfill\bm\square$

The above proof is identical to he complex result, the only difference is the definition of the rotation angle associated to either to Proposition \ref{rnP01} or the inner product (\ref{hn201}).
Thus we can follow the analogy with another result.

\begin{theorem}\label{HT01}  Let  $\,q(\alpha,\,t):J\times I\to\mathbbm H\,$ be a continuous deformation of the regular curve $\,q(t):I\to\mathbbm H,\,$ where  $\, J=[0,\,\infty)$. If $\,p(\alpha):J\times I\to \mathbbm H\,$  is a regular a continuous curve such that $\;p(\alpha,\,t)\cap q(\alpha,\,t)= \varnothing,\,$ the winding number $\mathcal R(q,\,p)$ is a constant independent of $\alpha$.
\end{theorem}
{\bf Proof:} From Proposition \ref{rnP03} and Lemma \ref{HP01},  the winding number 
\[ \mathcal R\big(q(\alpha,\,b),\,p(\alpha,\,b)\big)=\frac{1}{2\pi}\theta(\alpha,\,b) \] 
is a continuous function of $\alpha$. However, the winding number assumes only discrete values, and thus it is a constant independent of $\alpha$.

$\hfill\bm\square$

The above theorem is also valid in the case of symplectic curves, as well.  An immediate consequence of the Proposition \ref{rnP03} and Theorem \ref{HT01} is

\begin{corollary} Let  $\,q(\alpha,\,t):J\times I\to\mathbbm H\,$ be a continuous deformation of the regular curve $\,q(t):I\to\mathbbm H,\,$ where  $\, J=[0,\,\infty)$. 
Let $\,p(\alpha):J\to \mathbbm H\,$  be a continuous curve such that $\;p(\alpha,\,t)\cap q(\alpha,\,t)= \varnothing,\,$ and $\,\lim_{\alpha\to\infty}\big|p(\alpha)\big|=\infty.\,$ Thus 
\[\mathcal R\big(q,\,p(0)\big)=0.\] 
\end{corollary} 
The deformation of complex curves can be immediately adopted to quaternionic curves from Definition \ref{HD01}, and we can define:
\begin{definition}[Homotopy] 
Let $\,p,\,q:I\to\mathbbm H\,$ be two closed quaternionic curves. If the regular application $\,h(\alpha,\,t):[0,\,1]\times I\to\mathbbm H\,$ is a closed curve for every value of $\,\alpha\in J\,$ such that $\,h(0,\,t)=p(t)\,$ and $\,h(1,\,t)=q(t),\,$ is called an homotophy between $p$ and $q$, which are then called homotopic.
\end{definition}

Another immediate consequence of Theorem \ref{HT01} is

\begin{corollary}\label{HC01} 
Let $p,\,q:I\to U$ be two closed regular homotopic quaternionic curves and $\,U\subset\mathbbm H.\,$ If the regular curve $p_0(t)\notin U$, thus $\mathcal R(p,\,p_0)=\mathcal R(q,\,p_0)$. 
\end{corollary}

The winding number and the homotopy are related in the following theorems, that are analogous to plane curves.

\begin{theorem}[Quaternionic Poincar\'e-Bohl theorem] 
Let $p_0:I\to\mathbbm H$ be a regular quaternionic curve, and the regular closed curves $\,p,\,q:I\to \mathbbm H\,\backslash\,\{p_0(t)\}.\,$ If $\,p_0(t)\,$ does not intercept any segment that connects $\,p(t)\,$ and $\,q(t)\;\forall t\in I,\,$ then, $\,\mathcal R(p,\,p_0)=\mathcal R(q,\,p_0).\,$
\end{theorem}
{\bf Proof:} Building the homotopy $h=\alpha q+(1-\alpha)p$ and evoking Corollary \ref{HC01}, the theorem is proven.

$\hfill\bm\square$

Immediately, we have,
\begin{corollary}\label{HC03}
Let $p,\,q:I\to U$ be two closed regular homotopic quaternionic curves and $\,U\subset\mathbbm H.\,$ If $\,|p-q|<|q-p_0|,\,$ and $\,|p-q|<|p-p_0|\,$ hold $\,\forall\,t\in I\,$, then $\mathcal R(p,\,p_0)=\mathcal R(q,\,p_0)$.
\end{corollary}

Finally, we may obtain two interesting results that are analogous of complex analysis.

\begin{theorem}\label{HT02}
Let $\omega(t),\,p_0(t):I\to\mathbbm H$ be regular quaternionic functions. For each particular value of $t\in I,$ let $D_{\mathbbm H}$ be a disc of radius $R$, center $p_0,\,$ and border $\,C_{\mathbbm H}\,$ in the $\,\mathbbm R\times\mathbbm R \omega\,$ plane and let $F:D_{\mathbbm H}\,\to\,\mathbbm R\times\mathbbm R \omega\,$ be a continuous function. If the winding number is such that $\,\mathcal R\big(F(S_{\mathbbm H}),\,p_0\big)\neq 0,\,$ there is a point $\,q_0\in D_{\mathbbm H}\,$ such that $F(q_0)=p_0$. 
\end{theorem}
{\bf Proof:}  Let $s(\alpha,\,t)\,=\,p_0(t)\,+\,\alpha r\big(\cos(2n\pi t)\,+\,\omega\sin(2n\pi t)\big)$ be a circle in $\,\mathbbm R\times\mathbbm R\omega,\,$  where $n\in\mathbbm Z$ and of course $\,\mathcal R(s,\,p_0)=n.\,$ 
Let us suppose that $p_0\notin F(D_{\mathbbm H})$. Thus, he function $h(\alpha,\,t)=F(s(\alpha,\,t)-p_0)\,$ is an homotopy between $F$ and the constant $F(p_0)\,$ such that every curve belongs to a subspace of $\mathbbm H-\{p_0\}$.  Consequently, $F(S_{\mathbbm H})$ is homotopic to a point, whose winding number is zero and consequently 
$\mathcal R(F(S_{\mathbbm H}),\,p_0)=\mathcal R(F(S_{\mathbbm H}),\,p_0)= 0$. This contradicts the hypothesis, and thus there is a point $q_0$ such that $F(q_0)=p_0$. In particular, we can choose $p_0=$ and obtain $F(q_0)=0$.

$\hfill\bm\square$

Finally, we have:

\begin{theorem}[Fundamental theorem of algebra for quaternions] 
The quaternionic polynomial
\[
F(q)=\sum_{\ell=0}^n a_{n-\ell}\big(q-q_0\big)^\ell=0,\qquad\mbox{where}\qquad a_\ell\in\mathbbm R\qquad\mbox{and}\qquad q_0\in\mathbbm H,
\]
admits a quaternionic root.  
\end{theorem}
{\bf Proof:} We restrict $F(q)$  to the circumference $C_{q_0}$ of  center $q_0$ of radius $R=1+\sum_{\ell=0}^n|a_\ell|$. Using $q-q_0=R\big(\cos(2\pi t) +\omega\sin(2\pi t)\big)$ and $\,t\in[0,\,1]\,$ the winding number of the highest order term of the polynomial is $\mathcal R(a_n (q-q_0)^n,\,q_0)=n.$ The intent is to prove  that the entire polynomial has a nonzero winding number and then use Theorem \ref{HT02}. Therefore, over $C_{q_0}$, we have
\begin{align}
 \big| F(q)-a_n\big(q-q_0\big)^n\big|_{C_{q_0}} 
&=\left|\sum_{\ell=0}^{n-1} a_{n-\ell}\big(q-q_0\big)^\ell\right|_{C_{q_0}}\\
&\leq\sum_{\ell=0}^{n-1} \big|a_{n-\ell}\big| \left|\big(q-q_0\big)^\ell\right|_{C_{q_0}}\\
& <R^n =a^n\big|a_n(q-q_0)^n\big|_{C_{q_0}}
\end{align}
Using Corolary \ref{HC03}, we obtain  $\mathcal R(a_n (q-q_0)^n,\,q_0)=\mathcal R(F(C_{q_0}),\,q_0)=n.$ From Theorem \ref{HT02}, the polynomial has a root.

$\hfill\bm\square$

The fundamental theorem of algebra is already known for quaternions \cite{Niven:1944fta,Zhang;1997qmq}, but the above result is important because of their power of clarify the extension of the results of the article, whose range encompasses real coefficients polynomials only, and the main idea is simple, the quaternions are obtained using the generalization $i\to\omega(t)$, a transformation that does not change the orthogonal character between the real and the imaginary components. This simple idea, although it is not obvious, is the responsible to the novelty of the results contained in	 this article.

\section{CONCLUSION}

In this article we presended how to obtain the angle function for quaternionic curves, where the most important point is to choose a reference curve that permit us to express the quaternionic equations in terms of an unique imaginary unit. Thus, we obtained the winding number and the homotopy concept. 
The results generalize the two dimensional complex case, and we expect that this formulation will enable to shed brighter light over the previous results of quaternionic curves, and expanded a previous result \cite{Giardino:2021onv}. We expect that these results will be applied in physical theories, where quaternionic formalism is currently being deployed \cite{Giardino:2018rhs,Giardino:2021lov,Giardino:2020uab}. 

%
%
%
%

\begin{footnotesize}

\end{footnotesize}
\end{document}